\documentclass[twocolumn,aps,prl,showpacs]{revtex4}

\usepackage{epsf}
\usepackage{bm}

\begin{document}
    
\title{%
Why do gallium clusters have a higher melting point
than the bulk?}

\author{S. Chacko, Kavita Joshi, and D. G. Kanhere}

\affiliation{%
Department of Physics, University of Pune, 
Ganeshkhind, Pune 411 007, India}

\author{S. A. Blundell}

\affiliation{%
D\'epartement de Recherche Fondamentale sur la 
Mati\`ere Condens\'ee, CEA-Grenoble/DSM \\
17 rue des Martyrs, F-38054 Grenoble Cedex 9, France}

\date{\today}

\begin{abstract} 

Density functional molecular dynamical simulations have
been performed on Ga$_{17}$ and Ga$_{13}$ clusters to
understand the recently observed higher-than-bulk
melting temperatures in small gallium clusters [Breaux
{\em et al.}, Phys.\ Rev.\ Lett.\ {\bf 91},
215508 (2003)].
The specific-heat curve, calculated with the
multiple-histogram technique, shows the melting
temperature to be well above the bulk melting point of
303~K, viz.\ around 650~K and 1400~K for Ga$_{17}$ and
Ga$_{13}$, respectively.
The higher-than-bulk melting temperatures are
attributed mainly to the covalent bonding in these
clusters, in contrast with the covalent-metallic
bonding in the bulk.

\end{abstract}


\maketitle

Probing finite-temperature properties of clusters is a challenging
task, both experimentally and theoretically.  During the last few
years a number of intriguing aspects of the melting properties of
these finite-sized systems have been observed, such as a negative
microcanonical specific heat \cite{Na147} and substantially higher
melting temperatures than the bulk systems \cite{tinexpt}.  In a
series of experiments on free sodium clusters (with sizes ranging from
55 to 357 atoms), Haberland and co-workers \cite{Hab-CRP} observed a
substantial lowering (by about 30\%) of melting temperatures compared
to bulk sodium, with rather large size-dependent fluctuations in the
melting temperatures.  This lowering is in qualitative agreement with
old thermodynamic arguments \cite{thermo} that a small particle should
melt at a lower temperature than the bulk because of the effect of the
surface.  However, Jarrold and co-workers \cite{tinexpt} showed
recently that small tin clusters (with sizes between 10 and 30 atoms)
do not melt at least 50~K above the bulk melting temperature. 
Surprisingly, a very recent measurement on small gallium clusters by
Breaux {\it et al.}\
\cite{gaexpt-breaux} presented another example of a higher-than-bulk
melting temperature.  Their measurements indicated that
Ga$_{17}{}^{+}$ does not melt up to about 700~K, while Ga$_{39}{}^{+}$
and Ga$_{40}{}^{+}$ have melting temperatures of about 550~K, all well
above the bulk melting point [$T_m({\rm bulk}) = 303$~K].
While it might have been thought that tin was an exceptional case,
these new measurements suggest that elevated melting temperatures could
perhaps be a more widespread phenomenon.
These authors also investigated the
fragmentation pattern, but were unable to find the existence of any
particular stable building block, and did not offer any explanation
for the high melting temperatures.

Traditionally, classical molecular dynamics (MD) simulations have been
used to understand the finite-temperature behavior of
clusters~\cite{tradi-berry}.
For example, almost all simulations carried out to explain the thermodynamic
data of sodium clusters are based on classical MD employing a variety of
parametrized interatomic potentials \cite{NaCMD}.  However, these
attempts have fallen short of reproducing the crucial characteristics
observed experimentally, such as the precise sizes at which maxima in
the melting temperatures occur.  Quite clearly, for reproducing the
experimental results a more realistic treatment of interacting
electrons is desirable.  Recently, we have successfully demonstrated
the power of density functional molecular dynamics (DFMD) by providing
a detailed explanation of the experimentally observed phenomena of
higher-than-bulk melting temperatures for tin clusters \cite{sn10}. 
In particular, we have shown that the covalent bonding along with the
existence of a highly stable tricapped trigonal prism (TTP) subunit is
responsible for the higher melting temperature.  For gallium clusters
also, an {\it ab initio} treatment is quite crucial for simulating the
finite-temperature behavior, especially since there is a possibility
of a change in the nature of the bonding.

In this Letter, we provide an explanation and insight into the
phenomenon of the higher-than-bulk melting temperature recently
observed in small gallium clusters \cite{gaexpt-breaux}.  To this end
we have carried out {\it ab initio} density functional simulations
over a wide range of temperatures for the neutral clusters Ga$_{17}$
and Ga$_{13}$.  We present the calculated specific heat obtained via a
multiple-histogram (MH) analysis \cite{MH}.  We also present a
detailed analysis of bonding in these clusters and contrast it with
that of bulk.  In an earlier density functional calculation by Jones,
the bonds in small Gallium clusters have been found to be shorter than
those between the lighter atoms in the same main group, i.e.\
Al~\cite{gaclus-jones}.  As we shall see, these clusters indeed melt
at a temperature substantially higher than $T_m({\rm bulk})$, mainly
due to the formation of covalent bonds.

Isokinetic Born-Oppenheimer MD simulations have been carried out using
ultrasoft pseudopotentials within the local density approximation, as
implemented in the \textsc{vasp} package
\cite{vasp}.
For Ga$_{17}$, the MD calculations were carried out for
23 temperatures, each of duration 75~ps, in the range of $150
\le T \le 1100$~K, which results in a total simulation
time of 1.65~ns.  The resulting trajectory data have been used to
compute the ionic specific heat by employing the MH method
\cite{MH,amv-review}.  For all the calculations, we have used only
$sp$ electrons (4$s^2$ and 4$p^1$) as valence electrons, taking 3$d$
as a part of the ionic core (represented by an ultrasoft
pseudopotential).  We have verified that the $d$ electrons do not
significantly affect the finite temperature behavior and equilibrium
geometries by recalculating the equilibrium structures, and performing
three runs at finite temperature around the melting region, with
the $d$ electrons treated explicitly as valence electrons.

\begin{figure}
\epsfxsize=3.0cm
\centerline{\epsfbox{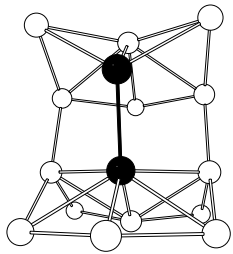}}
\caption{\label{fig.bulk}
A part of the bulk structure of $\alpha$-Ga (not a
unit cell).
It shows two buckled planes.
The dark line joining the black atoms corresponds to
the interplanar covalent bond discussed in the text.
}
\end{figure}

We begin our discussion by noting some interesting features of the
electronic structure of bulk Ga, which has been investigated by Gong
{\it et al.}\ \cite{gabulk-gong}.  The $\alpha$-Ga lattice, which is
stable at ambient pressure, can be viewed as base-centered
orthorhombic with eight atoms in the unit cell.
The peculiarity of this structure is that each atom has only one
nearest neighbor connected by a short bond at a distance of 2.44~{\AA}
(see Fig.~\ref{fig.bulk}).  The six other neighbors are at distances
of 2.71 and 2.79~\AA, in two sets of three.
These six atoms lie on strongly buckled parallel planes connected by
the short bonds, as shown in Fig.~\ref{fig.bulk}.  Electronic
structure calculations by Gong {\em et
al.}~\cite{gabulk-gong} reveal this short bond to be covalent in nature.
The density of states shows a pseudogap, which has been related to
this covalent bond \cite{Heine}, and the weak bonding in the buckled
planes leads to an observed metallic behavior.  Thus, two kinds of
bonds coexist in bulk Ga: one a molecular bond between the nearest
neighbors, and the other a metallic bond within the buckled planes.

\begin{figure}
\epsfxsize=6.5cm
\centerline{\epsfbox{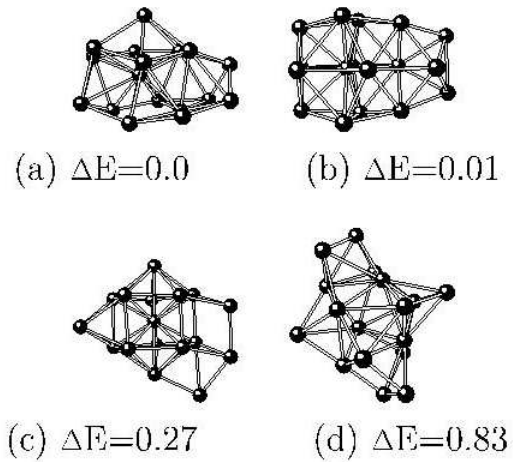}}
\caption{\label{fig.17str} 
   The ground state geometry and some of the the low
   energy structures of Ga$_{17}$.
   The structures are arranged in increasing order of
   the energy.
   Figure (a) represent the lowest energy structure and
   figure (d) represents the highest energy structure
   studied.  Energy differences $\Delta E$ are in eV.
   }
\end{figure}

Now we present and discuss some relevant features observed in the
equilibrium structures of Ga$_{17}$.  We have relaxed many structures,
randomly chosen from high-temperature DFMD runs.  In this way, we have
found more than 20 different equilibrium structures spanning an energy
range of about 0.83~eV with respect to the ground-state energy.  In
Fig.~\ref{fig.17str}, we show some low-lying structures relevant to
the present discussion.  A common feature observed in all these
geometries, except one (Fig.~\ref{fig.17str}-d), is the presence of a
trapped atom, that is, a single atom contained within a cage formed by
the remaining atoms.
The lowest-energy structure that we have found
(Fig.~\ref{fig.17str}-a) is a highly asymmetric structure, which can
be thought of as formed out of a decahedron with serious distortions
and asymmetric capping.  Interestingly, this structure can also be
viewed as composed of two near planar rings, as discussed further 
below.

\begin{figure}
\epsfxsize=7.5cm
\centerline{\epsfbox{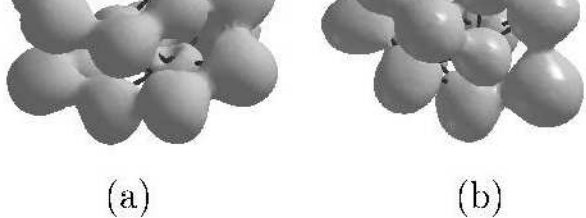}}
\caption{\label{fig.17iso}
    The isosurface of ELF for (a) Ga$_{17}$ at the
    value 0.68 and (b) Ga$_{17}{}^{+}$ at the value 0.65.}
\end{figure}

We have also analyzed the nature of bonding by employing the electron
localization function (ELF) \cite{elf-silvi}, which is defined such
that values approaching unity indicate a strong localization of the
valence electrons and covalent bonding.  The isosurface of the ELF for
the neutral Ga$_{17}$ and charged Ga$_{17}{}^{+}$ are shown in
Figs.~\ref{fig.17iso}-a and \ref{fig.17iso}-b, respectively.
A striking feature seen in these figures is the existence of strong
localized bonds giving rise to two ring-like structures.  Further,
these rings are also bonded to each other via the atoms at the edge of
each ring and with the trapped atom at the center.  The existence of
an isosurface of the ELF with such a high
value clearly indicates the covalent nature of the bonding.  This is
also substantiated by examination of the corresponding charge-density
isosurface (not shown).
Calculations of the ELF for the excited isomers in
Fig.~\ref{fig.17str} and for other small Ga clusters show similar
features.  Generally, most atoms have two covalent bonds in a roughly
linear arrangement, with a tendency to link neighbors together into
rings or ring fragments; extra covalent bonds on some atoms then join
these structures together.
The bonding here is thus seen to be in sharp contrast to the one
observed in the bulk $\alpha$-Ga, where, as discussed earlier, only a
single strong interlayer bond between two Ga atoms exists (see
Fig.~\ref{fig.bulk}).
An analysis of the bond lengths reveals that
each atom in the cluster has at least two nearest neighbors at a
distance of 2.5~{\AA} or less.
Indeed, it is this difference in bonding that is primarily responsible
for the higher melting temperature of the small clusters.

An examination of the molecular orbitals and the eigenvalue spectrum
(not shown) brings out some notable features.  The eigenvalue spectrum
is divided into two groups, which are separated by about 1.18~eV. The
lower 15 states forming the first group are the bonding states formed
out of atomic 4$s$ orbitals.  Almost all the upper states are formed
out of pure atomic $p$ orbitals and show no $sp$ hybridization.  The
only exception to this is the bonding between the trapped atom and its
three nearest neighbors, where a weak $sp^2$ hybridization is
observed.  This picture is also confirmed by a site-projected
spherical harmonics analysis of the orbitals, which does not show any
significant mixing of $s$ and $p$ character in these states.  All the
bonds seen are of predominantly $\sigma$ type, formed out of atomic
$s$ and $p$ orbitals.
The lack of $sp$ hybridization explains the tendency to form covalent
bonds at approximately 180$^{\circ}$ or 90$^{\circ}$ angles, as
observed in Fig.~\ref{fig.17iso}.

\begin{figure}
\epsfxsize=6.0cm
\centerline{\epsfbox{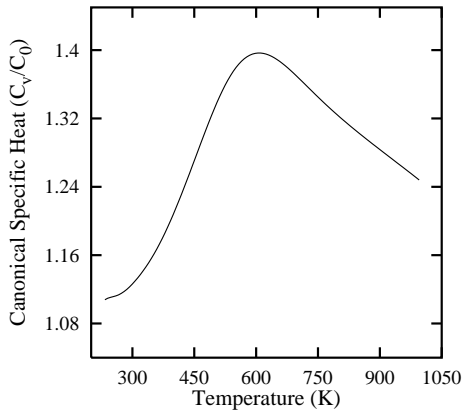}}
\caption{\label{fig.17cv}
    Normalized canonical specific heat of Ga$_{17}$
    cluster.
    $C_0=(3N-9/2)k_B$ is the zero temperature classical
    limit of the rotational plus vibrational canonical
    specific heat.
    }
\end{figure}

\begin{figure}
\epsfxsize=5.5cm
\centerline{\epsfbox{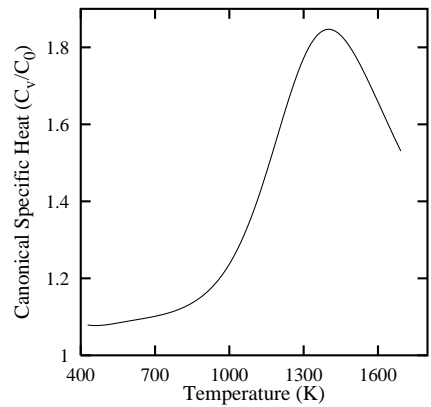}}
\caption{\label{fig.13cv}
    Normalized canonical specific heat of Ga$_{13}$
    cluster.
    }
\end{figure}               

The calculated specific heat curve for Ga$_{17}$ is shown
in Fig.~\ref{fig.17cv}.  A clear peak is observed in the specific heat
with a maximum around 650~K, well above the bulk melting point of
303~K. Following the discussion in Ref.~\cite{sn10}, we expect the
statistical uncertainty in our peak position to be up to 15\%.  Now,
a novel multicollision induced dissociation scheme has recently been
used to measure the caloric curve of small, charged Ga clusters
\cite{gaexpt-breaux}.  For Ga$_{17}{}^{+}$, no
evidence was found for melting (in the sense of a peak in the specific
heat) over a temperature range 90--720~K. Our simulations are
consistent with this finding.  Note that there is likely to be some
shift of the melting temperature between the neutral and charged cluster.

While the thermodynamic simulation has been carried out for neutral
Ga$_{17}$, calculations of the ELF in Ga$_{17}{}^{+}$ reveal a similar
network of covalent bonds (see Fig.~\ref{fig.17iso}b).
The ground-state geometry changes to a
more symmetric form (which is a low-lying isomer for Ga$_{17}$), and
the HOMO-LUMO gap increases from 0.76~eV for the neutral to 1.07~eV.
The low-lying isomers span an energy range of order 0.8~eV for both
Ga$_{17}$ and Ga$_{17}{}^{+}$.  Therefore, we expect Ga$_{17}{}^{+}$
to have similar melting properties to Ga$_{17}$.  The larger HOMO-LUMO
gap for Ga$_{17}{}^{+}$ may imply that it has a slightly higher melting
temperature.

Our specific-heat curve shows no premelting features in the
temperature range studied, and that the peak is quite broad, as
expected for the finite-sized systems.  A detailed examination of the
ionic motion at different temperatures sheds more light on the nature
of melting.  Around 225~K, we observe the occurrence of the first
excited state (Fig.~\ref{fig.17str}-b).  As the temperature rises
further, the system visits a variety of isomers all of which contain a
trapped atom.  It is only above 550~K that the isomer shown in
Fig.~\ref{fig.17str}-d is observed.  Evidently, the peak in the
specific heat is associated with the diffusion of the trapped atom out
of the enclosing structure.  Interestingly, in the trajectories at all
temperatures below 600~K, the trapped atom is seen to bond itself to
atoms in both rings.  This confers stability on the caged structure.  We
have also carried out the analysis of the melting properties via
traditional parameters such as the root-mean-square bond-length
fluctuations and the mean square atomic displacements (not shown). 
Their behavior is consistent with the above observations.

To summarize, our {\it ab initio} MD simulations have shown that the
elevated melting temperature of Ga$_{17}$ is due mainly to the
formation of covalent bonds.  The stability of the cluster is further
enhanced by the role of the caged atom.  These observations should
have implications for the melting characteristics of clusters of
smaller sizes, such as Ga$_{13}$.  The ground-state geometry of
Ga$_{13}$ is decahedral, a more compact structure having a greater
average number of covalent bonds per atom than Ga$_{17}$.
Further, our bonding analysis shows that the bonding is
strongly covalent, similar to Ga$_{17}$.  We therefore expect this
cluster to melt at a higher temperature than Ga$_{17}$.  To verify
this conjecture, we have carried out extensive DFMD simulations for 30
different temperatures in the range $40 \le T \le 1750$~K, with a
total simulation time of about 2.7~ns.  The resulting specific heat is
shown in Fig.~\ref{fig.17cv}.  Indeed, the peak is around 1400~K, a
much higher value than the peak position of Ga$_{17}$ or the bulk
melting point.  The detailed analysis will be published elsewhere.

\begin{figure}
\epsfxsize=5.0cm
\centerline{\epsfbox{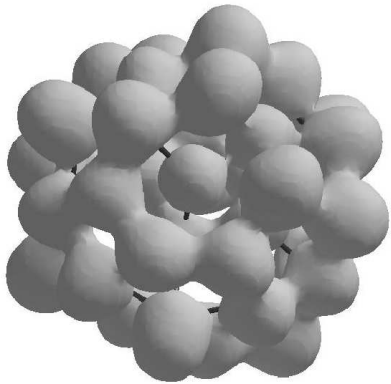}}
\caption{\label{fig.40elf}
    The isosurface of ELF for Ga$_{40}$ at the value 0.64.
    }
\end{figure}               

Finally, we note that Ga$_{39}{}^{+}$ and Ga$_{40}{}^{+}$ have been
measured to have a broad maximum in the specific heat around 550~K
\cite{gaexpt-breaux}.  Unfortunately, the large sizes of these 
clusters prevent us from making a complete thermodynamic study at this
stage.  However, we have found and analyzed some low-lying equilibrium
structures of Ga$_{40}$.  We believe that a mechanism similar to that
in Ga$_{17}$ is operative in Ga$_{39}{}^{+}$ and Ga$_{40}{}^{+}$.  In
Fig.~\ref{fig.40elf} we show an isosurface of the ELF for the value of
0.64 for Ga$_{40}$.  The ring-like structures of covalent bonds are
once again clearly evident.  However, it may be noted that at this
value of the ELF no contours are seen between the inner core atoms and
the outer surface, and not all the atoms on the surface are bonded. 
Therefore, it is not unreasonable to expect this cluster to melt at a
somewhat lower temperature than Ga$_{17}$, but still higher than that
of the bulk.

In conclusion, we have carried out extensive density functional
thermodynamical simulations on Ga$_{17}$ and Ga$_{13}$ with the aim of
understanding the observed higher-than-bulk melting temperatures.  The
analysis of the specific-heat curve indicates the melting temperatures
(defined as the peak of the specific heat curve) to be around 650 and
1400~K for Ga$_{17}$ and Ga$_{13}$, respectively.  This result is
consistent with the recent experimental observations of Breaux {\it et
al.}\ \cite{gaexpt-breaux}.  We find a significant change in the
nature of bonding between bulk Ga and small Ga clusters.  The strong
covalent bonds in the small clusters, along with the stabilizing role
of the trapped atom in Ga$_{17}$, are responsible for the
higher-than-bulk melting temperatures.  While the structural details
for Sn and Ga clusters are very different, the common feature between
these two systems is thus seen to be the development of strong
covalent bonding in the small clusters.

One of us (SC) acknowledges financial support from CSIR (New Delhi). 
It is a pleasure to acknowledge C-DAC (Pune) for providing us with
supercomputing facilities.

\end{document}